\documentclass[sn-basic]{sn-jnl}
\usepackage[T1]{fontenc}
\usepackage[utf8]{inputenc}
\usepackage{natbib,aas_macros}

\jyear{2022}%

\theoremstyle{thmstyleone}%
\theoremstyle{thmstyletwo}%

\theoremstyle{thmstylethree}%

\begin{document}

\title[Dusty plasma in active galactic 
nuclei]{Dusty plasma in active galactic 
nuclei}


\author*[1]{\fnm{Bo\. zena} \sur{Czerny}}\email{bcz@cft.edu.pl}

\author[2]{\fnm{Michal} \sur{Zaja\v{c}ek}}\email{zajacek@mail.muni.cz}

\author[1,3]{\fnm{Mohammad-Hassan} \sur{Naddaf}}\email{naddaf@cft.edu.pl}
\author[1,3,4]{\fnm{Marzena} \sur{Sniegowska}}\email{msniegowska@tauex.tau.ac.il}

\author[5]{\fnm{Swayamtrupta} \sur{Panda}}\email{spanda@lna.br}

\author[3]{\fnm{Agata} \sur{R\' o\. zanska}}\email{agata@camk.edu.pl}

\author[6]{\fnm{Tek P.} \sur{Adhikari}}\email{tek@iucaa.in}

\author[1]{\fnm{Ashwani} \sur{Pandey}}\email{ashwanitapan@gmail.com}

\author[1]{\fnm{Vikram Kumar} \sur{Jaiswal}}\email{vkj005@gmail.com}

\author[7]{\fnm{Vladim\'{\i}r} \sur{Karas}}\email{vladimir.karas@asu.cas.cz}

\author[7]{\fnm{Abhijeet} \sur{Borkar}}\email{abhijeet.borkar@asu.cas.cz}

\author[8]{\fnm{Mary Loli} \sur{Mart\'inez-Aldama}}\email{mmartinez@das.uchile.cl}

\author[1]{\fnm{Raj} \sur{Prince}}\email{raj@cft.edu.pl}


\affil[1]{\orgdiv{Center for Theoretical Physics}, \orgname{Polish Academy of Sciences}, \orgaddress{\street{Al.~Lotnik\' ow 32/46}, \city{Warsaw}, \postcode{02-668}, \country{Poland}}}

\affil[2]{\orgdiv{Department of Theoretical Physics and Astrophysics}, \orgname{Faculty of Science, Masaryk University}, \orgaddress{\street{Kotl{\'a}\v{r}sk{\'a} 2}, \city{Brno}, \postcode{611 37}, \country{Czech Republic}}}

\affil[3]{\orgdiv{Nicolaus Copernicus Astronomical Center}, \orgname{Polish Academy of Sciences}, \orgaddress{\street{Bartycka 18}, \city{Warsaw}, \postcode{00-716}, \country{Poland}}}

\affil[4]{\orgdiv{School of Physics and Astronomy}, \orgname{Tel Aviv University}, \city{Tel Aviv}, \postcode{69978}, \country{Israel}}

\affil[5]{\orgdiv{Laborat\'orio Nacional de Astrof\'isica}, \orgname{MCTI}, \orgaddress{\street{R. dos Estados Unidos 154}, \city{Itajub\'a}, \postcode{37504-364}, \country{Brazil}}}

\affil[6]{\orgname{Inter University Center for Astronomy and Astrophysics}, \orgaddress{\street{Maharashtra}, \city{Pune}, \postcode{411007}, \country{India}}}

\affil[7]{\orgdiv{Astronomical Institute}, \orgname{Czech Academy of Sciences}, \orgaddress{\street{Bo\v{c}n\'{\i}}~II~1401}, \city{Prague}, \postcode{14100},  \country{Czech Republic}}

\affil[8]{\orgdiv{Departamento de Astronom\'ia}, \orgname{Universidad de Chile}, \orgaddress{\street{Camino del Observatorio 1515}, \city{Santiago}, \postcode{Casilla 36-D Correo Central}, \country{Chile}}}



\abstract{For many years we have known that dust in the form of a dusty-molecular torus is responsible for the obscuration in active galactic nuclei (AGN) at large viewing angles and, thus, for the widely used phenomenological classification of AGN. Recently, we gained new observational and theoretical insights into the geometry of the torus region and the role of dust in the dynamics of emerging outflows and failed winds. We will briefly touch on all these aspects, and provide a more detailed update of our dust-based model (FRADO -- Failed Radiatively Accelerated Dusty Outflow) capable of explaining the processes of formation of Balmer lines in AGN.}

\keywords{galaxies: active, quasars: emission lines, supermassive black holes, radiative transfer}



\maketitle

\section{Introduction}\label{sec1}

Dust is present in all galaxies, both active and non-active, as it is one of the constituents of the interstellar medium \citep[see][for a review]{draine2003}. In galaxies, dust originates primarily from vigorous, stellar winds characteristic of evolved stars which enriched their chemical composition through nuclear burning. Supernovae explosions are notably efficient in ejecting a highly enriched 
material. Our Galaxy -- Milky Way -- also contains a considerable amount of dust \citep{calzetti2000,weingartner2001}. 

We are located at a distance of about $8$ kpc from the Galactic center, roughly in the Galactic plane, and so the nucleus is shielded from us by an enormous amount of extinction of the order of 25 mags in V-band \citep[e.g.][]{Nishi2008, Schodel2010,noguerras2020}. However, the Milky Way nucleus is not active at present stage, and its central black hole is currently dim and inactive. In active galactic nuclei (AGN), the dust located at inner $\sim 10$ pc plays a very specific role -- which is directly associated with their enhanced accretion activity.

 In this brief review, we attempt to summarize a broad range of topics related to the role of dust in AGN classification, the geometry of the dust distribution, the physical properties of the dust and its interaction with the gaseous medium, and the dynamics of the dusty medium. We summarize the unsolved problems and prospects in Section~\ref{sect:summary}. Our review is far from complete, although we include several references to other works that are particularly relevant to the topic.

\section{Basic properties of AGN and the dusty torus in the standard unification scheme}
\label{sect:basic}

Over the fifty years of the studies of quasars and even longer investigation of other types of active galaxies, such as Seyfert galaxies, radio galaxies and blazars, we have achieved a basic understanding of the structure of these objects 
(see monographs by \citealt{krolik_book,Meier_book}, and reviews by \citealt{netzer2015,bianchi2022} for a recent comprehensive overview of AGN properties). Every AGN consists of a supermassive black hole, 
and an inflowing material which determines the level of activity. The inflowing material usually possesses considerable angular momentum; close to the nucleus it must form a disk-like structure. In highly accreting sources, the plasma cools efficiently, and a standard accretion disk forms, roughly represented by the model of \citet{SS1973}. If the stream of material is less vigorous, the flow becomes hot and optically thin, at least in the innermost parts. Advection-Dominated Accretion Flow (ADAF) then represents a canonical example \citep{narayan1994}. The actual geometry must be more complicated to explain the hard X-ray emission and the soft X-ray excess on top of the optical/UV Big Blue Bump coming from the cold disk \citep{czernyElvis1987} \citep[for reviews and specific models, see e.g.][]{abramowicz1995,beloborodov1999,done2012}. The character of the flow in the outer parts is still under discussion and may have a character of the spherically symmetric  flow \citep{bondi1952} or disk-like shape, depending on the feeding mechanism, the role of stellar winds, etc. \citep[see e.g.][for a review]{thaisa2008}.

Some AGN (about 10\%) exhibit strong relativistic outflows in the form of jets, and these are classified traditionally as radio-loud objects or (more recently) as jetted objects 
\citep[see e.g.][for a review]{Padovani2017}. Out of those objects, some have jets oriented close to the line of sight
towards an observer, so the jet emission is boosted and frequently overshines the nuclear emission. 
In the later sections, we will concentrate on the radio-quiet (non-jetted) sources, where we have much better insight into
processes going on near the black hole. In the case of radio-quiet sources, we follow processes which are sensitive to the flow description close to the event horizon (e.g., cold disk emission in UV, X-ray reflection spectrum in X-rays, including the iron K$\alpha$ line, X-ray reverberation mapping etc.) while in jetted sources the direct resolution gives insight only on the sub-parsec scale. 

Radio-quiet sources (about 30\% of them, \citealt{veron_cetty2010,shen2011,rakshit2020}) frequently show broad and narrow emission lines. The line broadening must be caused by the dispersion in the velocity of the emitting material since the kinematic widths of narrow lines are of the order of a few hundred km s$^{-1}$, and the broad lines have line widths of the order of a few thousand km s$^{-1}$, while the expected thermal broadening is of the order of a few km s$^{-1}$ \citep{kollatschny2011}. The latter constraint comes from the fact that Balmer lines are intense, implying that the hydrogen is not fully ionized and the corresponding plasma temperature must be of the order of 10 000 K. Much higher temperature would lead to full ionization of hydrogen and Balmer lines would not be observed \citep{collin1987}. Therefore, the standard view is such that the emission comes mainly from clouds orbiting the central black hole at a distance of a fraction of a parsec (for broad lines) and above a parsec (for narrow lines). These regions are named the Broad Lines Region (BLR) and the Narrow Line Region (NLR), respectively. The line emission is caused by the irradiation of plasma by the central region where most of the accretion energy is dissipated and re-emitted \citep{wills1985,osterbrock2006}. When broad lines are seen, narrow lines are also present, and this shows up as a complex line shape of a single line consisting of broad and narrow components.

The emission lines differ not only in emitted velocities but also in the mechanism of their formation. The Balmer lines are seen both as broad and narrow lines in a given source but some of the narrow lines coming from other elements are actually forbidden lines, such as the [OIII] line at $\lambda5007$ \AA, which means that the density of the emitting material is much lower in the NLR \citep[e.g.][]{baskin2005,bennert2006,MS2011,kakkad2018}. A broad (1000--3000 km s$^{-1}$) [OIII]$\lambda5007$ component has been also detected in some sources with high accretion rates. This component is mostly observed with a blue-shifted profile, which indicates that part of the [OIII]$\lambda5007$ is originates in an outflow \citep{zamanov2002, leipski2006, negrete2018, schmidt2018, sexton2021}. The line width, in such cases, does not represent the Keplerian virialized motion but instead the accelerated outflow. Some sources also show a broad [OIII]$\lambda$5007 core at the rest frame or in redshifted wings \citep{negrete2018,berton2021}. 

A fraction of the sources shows only narrow lines. It was a puzzle for many years about why BLR is not visible in these objects. Most quasars show broad lines, but fainter Seyfert galaxies frequently did not show BLR, so Seyfert galaxies were further sub-classified as Seyfert 1 (with BLR) and Seyfert 2 (without BLR). Quasars were also divided into type 1 and type 2 by analogy, although type 2 is rare.  The resolution came
from polarimetric observations performed by \citet{antonucci1985} for the famous Seyfert galaxy NGC 1068 from the original list of 
active galaxies selected by Seyfert in his seminal paper \citep{seyfert1943}. When viewed under unpolarized light, the source did not show the BLR component, while, in the polarized light, the broad component of the Balmer H$\beta$ line was clearly present. The interpretation was that a thick torus surrounds the nucleus, and the BLR is located closer than the torus, while the NLR is located further out. If the viewing angle is low (as measured with respect to the symmetry axis), the view of the nucleus is unobscured, and we can see both NLR and BLR line components, but if the viewing angle is
high, the torus shields the BLR component. The BLR component is only revealed again in the polarized light since, in this case, we can 
detect only a small fraction of the total light that has been scattered by the low-density material filling the (almost empty) space around the torus. This torus is now known as a dusty/molecular torus, with an inner radius of about a parsec \citep{koshida2014}. The torus itself is seen in the infrared part of the AGN spectrum. Since the dust sublimation temperature is of the order of 1500 K, the dust emission appears only at longer wavelengths. Dust cannot be much hotter -- it then evaporates and the material is changed into (rich in heavy elements) gaseous phase \citep{barvainis1987,kishimoto2007}. Further insight came from observing and modelling this source \citep[e.g.][]{miller1991,Seba2006,garcia2014}. The location of the dust is observationally very well constrained by the reverberation mapping \citep[see][for recent results]{minezaki2019}.

Thus, the dusty/molecular torus provided a key element in the unification picture of radio-quiet AGN. It implies that their structure
is always the same; the apparent variety of types comes from different orientations. The truth is slightly more complex: the accretion rate onto the black hole is another key element. It is conveniently parameterized by the Eddington ratio, i.e., the ratio of the luminosity dissipated in the flow to the maximum luminosity of a spherical source in which the radiation pressure (calculated for fully ionized pure hydrogen) marginally balances the gravity of the central body. This ratio does not depend on the value of the black hole mass, which in AGN can range from a few hundred thousand solar masses to ten billion solar masses. Hence, it is a very convenient value to use, even if, with limitations; an AGN is neither spherically symmetric nor consists of pure hydrogen. In quasars, the Eddington rate is typically about 10 \% (between 1 \% and 100 \%, see e.g. \citealt{panda2018}, their Fig.~3, \citealt{MaryLoli2020}, their Fig.~3). In nearby AGN, it is frequently much lower. Thus, quasars seem to always have a torus, while for nearby sources, there may be
a considerable fraction of true Seyfert 2 galaxies where there is no torus. 

Spectra of faint sources are strongly dominated by starlight, so the corresponding polarization measurements are difficult, and broad lines are also difficult to distinguish against the strong starlight background. On one hand, extremely inactive galaxies like the center of the Milky Way -- Sgr A* -- do not have a dusty torus (its Eddington ratio is about $10^{-9}$; see e.g. \citealt{EHT_Sgr_2022}, \citealt{2017FoPh...47..553E}), but they also are most likely devoid of any BLR itself since the inner flow is optically thin (e.g. ADAF), and, the BLR clouds do not form \citep{laor2003,nicastro2003,czernyKurasz2004}. On the other hand, careful observation of potentially true Seyfert 2 galaxy NGC 3147  with the Hubble Space Telescope revealed the existence of broad H$\alpha$ line \citep{bianchi2019,bianchi2022a} although the Eddington ratio in this source is quite low, about $10^{-4}$; in this case, the line was not shielded, instead, it was just difficult to detect. The sources are divided into HBLR (Hidden BLRL) sources, where the broad lines were finally detected, usually in the polarized light, and NHBLR (Non-Hidden) BLR, or bare Seyfert 2 galaxies. With advancing observational accuracy, some sources are reclassified as HBLR \citep[e.g.][]{ramosAlmeida2016}. The physical conditions (e.g., the limit for the Eddington ratio) needed for the absence of the BLR and the torus are not yet well specified. It remains to be seen if the true absence of the BLR is related to the absence of also the torus, which would then suggest a common origin of these two important elements of an AGN and (most probably) their relation to the cold standard accretion disk.

The Eddington rate also influences the shape of the broad-band spectrum that is related to the character of the inner flow close to the black hole. In higher Eddington sources (Eddington ratio above 1\%), the inflow is in the form of a cold, optically thick, geometrically thin accretion disk described rather well by the standard model of \citet{SS1973} and its relativistic version \citep{novikov1973}. As an alternative, the slim disk operates at very high Eddington rates \citep{abramowicz1988}. The spectrum is dominated by the Big Blue Bump component \citep[e.g.][]{cz_elvis_1987}. At lower Eddington rates, the inner flow is most likely optically thin, in the form of an advection-dominated accretion flow (ADAF; \citealt{ichimaru1977,narayan1994}). Various extensions of this scenario were proposed, particularly in the context of jetted sources, such as JED-SAD (jet emitting disk -- standard accretion disk) models \citep{ferreira1995,petrucci2008} or MAD (magnetically arrested disk) models \citep{MAD2003,narayan2022}. The unification scheme actually requires three parameters: jet strength, orientation, and the Eddington rate. Apart from the issue about the torus, the unification scheme works surprisingly well over a wide range of black hole mass, and there is also a good analogy in spectral behaviour between 
AGN and Galactic black holes of masses of the order of only 10 solar masses \citep[e.g.][]{sobolewska2011,moravec2022}.

All three above-mentioned parameters strongly influence the broad band spectrum of an AGN. The nucleus itself is basically unresolved, however, by observing the spectrum and the variability of the source the AGN structure can be constrained. Only two nearby objects have been resolved so far: the radio galaxy M87 and the center of the Milky Way -- Sgr A* -- were resolved by the Event Horizon Telescope to such an extent that the shadow, or silhouette, of the central black hole, is revealed and the innermost pattern of the hot flow in semi-circular motion \citep{EHT_M87_2019,EHT_Sgr_2022} is seen. These interferometrical observations were performed in the millimeter regime, optimized for the resolution and the extinction. The detected signal was the synchrotron emission from hot, low-density plasma. We note that both objects are examples of sources at very low Eddington rates, so no cold disk was observed. 

\section{On the coexistence of the hot and cold dust-less plasma}
\label{sect:hot_warm}

The spectral complexity of AGN -- broad band continuum and various types of emission lines -- implies that the plasma in AGN has a broad parameter range for the density and ionization state. In addition, AGN shows a range of absorption lines that arise from the plasma located along the line of sight towards the black hole, where most of the optical/UV/X-ray emission is produced. These lines are phenomenologically characterized by their width, strength, and ionization of the corresponding ions. We thus have narrow absorption lines (NAL; e.g. \citealt{wampler1995,elvis2000}), broad absorption lines (BAL; e.g. \citealt{weyman1991,choi2022}), warm absorbers (WA; e.g. \citealt{nandra1994,holczer2007,adhikari2019}), and the ultra-fast outflows (UFO; e.g. \citealt{tombesi2010}). The distance of the material forming those lines is difficult to estimate because we cannot identify the line widths or shifts with Keplerian motion, the way we do it for the BLR. 

It may be that some of these absorption lines are actually due to BLR or NLR, seen in absorption instead of emission. Absorption lines imply a clumpy structure of the absorbing medium since numerous kinematic components are observed, particularly in UV, with different velocity shifts to the rest frame of an AGN. 

That scenario might pose a problem with cloud stability since a dense, non-self-gravitating cloud in space would expand on the dynamical timescale and will eventually get dispersed. However, if cooler, denser clouds are indeed embedded in much hotter, low-density, fully ionized plasma, they can be stabilized. This scenario is well justified theoretically. As pointed out by \citet{krolik1981} in the context of AGN, irradiated plasma is subject to thermal instability, and within a certain range of the gas pressure, two solutions for the gas temperature can co-exist -- a low-density hot plasma, at the temperature of the order of $10^7$ K, which heats and cools by the Compton process, and colder, denser plasma at the temperature of the order of $10^4$ K, which mostly heats and cools by atomic processes. In general, all radiative processes are important in partially ionized plasma, and they include Comptonization (heating and cooling), free-free transitions, bound-free as well as bound-bound transitions. Thus, plasma at two extremely different temperatures can coexist. The equilibrium curves are conveniently shown using the ionization parameter $\Xi$ introduced by \citet{krolik1981},
\begin{equation}
\Xi = \frac{F_{\rm ion}}{nkTc },
\end{equation}
where  $F_{\rm ion}$ is the ionizing flux between 1 and $10^3$ Rydberg, $n$ is the local number density, $T$ is the medium temperature, $k$ is the Boltzmann constant, and $c$ is the speed of light. This parameter is dimensionless, and it measures the ratio of the radiation pressure to gas pressure. 

The colder clouds can be in equilibrium when they are located on the equilibrium curve at an identical value of $\Xi$ on the lower and upper branches. The intermediate branch is unstable, and under small perturbations, the plasma rapidly evolves on the thermal timescale towards one of the stable branches.

The coexistence of the two plasma states with a sharp discontinuity is more complex. Clouds out of equilibrium can reach equilibrium, but they can also slowly grow through condensation or disappear by 
evaporation. Computations of clouds out of equilibrium require the inclusion of an additional process: electron conduction, which is important at the sharp border
between the two phases, where the temperature gradient is significantly large. This was recognized by \citet{field1965} and subsequently studied in several papers in the context of 
AGN, both for the transition between the colder disk and the hotter plasma at the disk corona and for the warm absorber (WA) and BLR clouds \citep[e.g.][]{begelman1990,rozanska1999,loic2007}. These computations require proper treatment of the radiative processes. Complex radiative transfer codes are used, such as CLOUDY \citep{CLOUDY2017}, which allow us to calculate the ionization state of the material, and its temperature and perform the radiative transfer.
The conduction term is frequently neglected if only the stationary solution is of interest, but then there are problems with the uniqueness of the solution under constant pressure. 

Numerous papers employ the equilibrium curve for the discussion of the properties of the multi-phase medium in the context of AGN. For example, \citet{pittard2001} used the equilibrium curves to address the cloud formation in the colliding AGN-supernovae winds. Several authors  \citep[e.g.][]{krolik2001, Gupta2013,adhikari2015,adhikari2019} considered the pressure balance in different phases of the WA medium. The issue is further complicated if we consider outflowing medium \citep{dannen2020,waters2021}, or even supersonic outflow velocities \citep{waters2022}.

The scenario of the coexistence of the colder clouds and the intercloud medium was also formulated as the radiation pressure confinement (RPC) of the clouds \citep[see][and the references therein]{stern2014,baskin2021}. In any case, the rough pressure balance and approximate thermal equilibrium allow for cloud formation and relatively long cloud existence. There are, however, mechanisms destroying such clouds \citep[see discussion in ][]{muller2022}, e.g., in cloud collisions, the action of tidal forces, Kelvin-Helmholtz instability, etc. The clumpy medium is always dynamic. The same is true for the interstellar medium, although timescales in AGN are much shorter due to the smaller sizes of the clouds and much higher local densities.

\section{Coexistence of the hot and cold dusty plasma}

While the conditions of coexistence of the hot $T \sim 10^7$ K and warm $\sim 10^4 $ K plasma was frequently discussed in the context of AGN (see Section~\ref{sect:hot_warm}), the coexistence of the third phase, cold dust $ T < 1500 $ K with the other two phases is less well studied from the physical point of view. On the other hand, we know that the dust is present within the central 1 pc as the dusty/molecular torus (see Section~\ref{sect:basic}). In the advanced torus models, the clumpiness of the medium is included \citep[e.g.][]{nenkova2008a,nenkova2008b,stalevski2012,siebenmorgen2015,2023arXiv230107735S} since otherwise the more distant parts of the torus are not adequately illuminated and the model does not reproduce the observed silicate features in absorption or emission. The inter-cloud medium is usually neglected, like in \citet{nenkova2008a,nenkova2008b}, \citet{stalevski2012}, and \citet{siebenmorgen2015}, who include the inter-cloud dust but not the inter-cloud gas phase. 

The first extremely influential paper which addressed the issue of the dusty plasma interaction with the incident radiation was published by \citet{netzer_laor_1993}. Their study of the dusty plasma and its efficiency in the production of emission lines showed that the presence of dust strongly suppressed the line emissivity, and, thus, they showed that the gap between the BLR and NLR (see Section~\ref{sect:basic}) is caused by the presence of the dust. This way one of the key puzzles in AGN was explained.

However, later, a new sub-type of AGN was introduced -- Narrow Line Seyfert 1 galaxies \citep{osterbrock1985}. These sources are different from type 2 sources, they exhibit broad lines, but much narrower than in typical Seyfert galaxies, and their forbidden lines are weak, unlike in Seyfert 2 objects. A formal definition included the maximum width of the lines of 2000 km s$^{-1}$. Later, for quasars which typically have much higher mass, a similar new class was introduced, i.e., type A (with the limit of the line width of 4000 km s$^{-1}$), while typical quasars were classified as type B, with broader lines \citep{sulentic2000}. Subsequent studies showed that these sources accrete close to the Eddington limit and, they form a high accretion rate tail of all AGN \citep{Panda_2019ApJ...882...79P, Panda_2022arXiv221015041P}. In addition, these objects do not show two-component line profiles, and their line profiles can be fitted by a single Lorentzian shape \citep[e.g.][]{sulentic2000,kollatschny2013,zajacek2020}. The explanation of this lack of the BLR -- NLR gap in these sources was solved by \citet{adhikari2016} where the authors showed that if the medium density is high enough, then the dust does not compete efficiently for photons with the partially ionized plasma. The gap postulated by \citet{netzer_laor_1993} does not form, and indeed objects accreting at a high Eddington rate have a higher local density in the BLR than less vigorously accreting sources \citep{panda2018}. 

\begin{figure}
\centering
\includegraphics[width=0.95\textwidth]{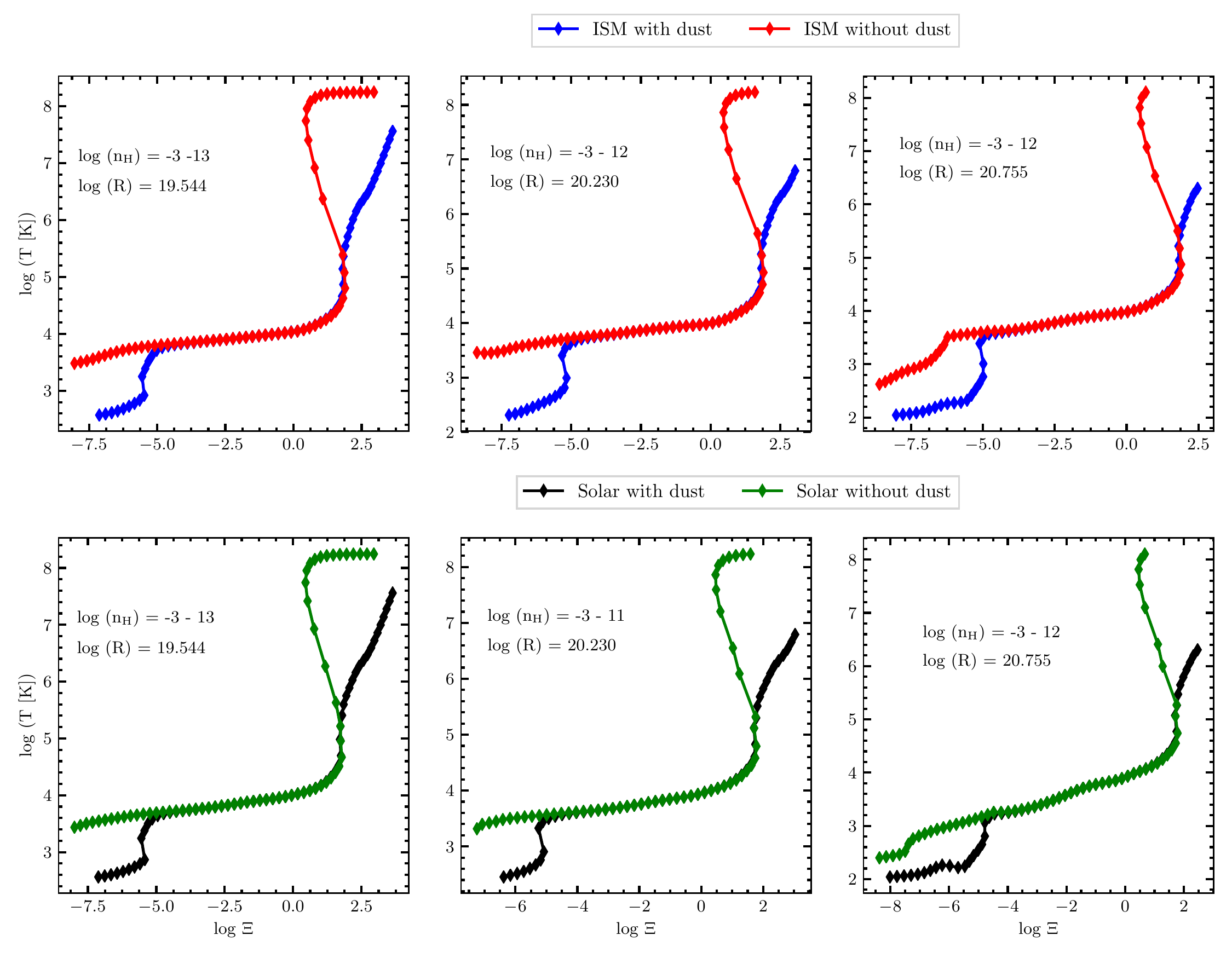}
\caption{\label{fig:borkar} Stability curves for the circum-nuclear medium for two different, heavy element abundances: upper panels show ISM abundances, while bottom panels show solar abundances, with
increasing radii from left to right: 11.5, 55, and 180 pc. Each panel contains calculations with and without dust. Each point in the curve represents the model computed for one starting value of density. The range density range is given in each panel. The SED luminosity is always kept at $L_{bol} =  10^{43}$ erg s$^{-1} $ (from \citealt{borkar2021}).}
\end{figure}

The coexistence of the three phases in AGN was finally studied in detail by \citet{borkar2021} using the CLOUDY code in the context of high-resolution observational data for Centaurus A from ALMA and Chandra. We constructed the equilibrium curves for dusty plasma irradiated by hard X-rays from the central region (see Figure~\ref{fig:borkar}). Here, the temperature $T$ is the plasma temperature. We notice that without dust, the temperature saturates just below $10^4$ K, even for a low value of the ionization parameter $\Xi$. Only at the distance of $\sim 200$ pc does the temperature drop below $10^3$ K. At the highest values of $\Xi$, the plasma temperature saturates at Inverse Compton temperature, about $10^8$ K for the adopted shape of the radiation spectrum. We see the unstable branch corresponding to the negative slope of the stability curve. This agrees with the possibility of the coexistence of warm and hot plasma. However, in the presence of the dust, the temperature drops to much lower values for high $\Xi$, forming, at this location, an additional branch of solutions with an instability present at $\Xi \sim 10^{-5}$. There we can have a coexistence of cold and warm plasma. 

For high $\Xi$, the temperature of the dusty plasma is much lower due to the interaction with dust grains. At $\Xi \sim 100$, there is a narrow range of the coexistence of the warm and hot plasma. Note that the temperature of the dusty plasma can be as high as $10^7$ K. The dust can still exist, and its temperature is lower than the sublimation temperature, as reported by the CLOUDY code, but the dust temperature is not plotted. Thus the dusty medium at high $\Xi$ is a two-temperature medium, with individual cold dust grains embedded in the hot plasma. This is possible due to the low density of the hot plasma.  

The computations thus show two ranges of $\Xi$ at which a clumpy medium can form, but all three phases cannot coexist. A cold, dusty cloud cannot be in pressure equilibrium with the hot plasma, at least in the presented computations that consider an optically thin medium. In the optically thick medium, such coexistence is, perhaps, possible but the code CLOUDY cannot be used to describe such a scenario. Computations performed by \citet{borkar2021} also did not address the processes taking place closer to the black hole within the BLR medium. 

In the computations we used the currently most representative broad band spectrum of Cen A to describe the incident radiation, as given in Fig.~4 of \citet{borkar2021}. The result may depend on this assumption, so it is not necessarily good for all sources. We also neglected the dust destruction mechanisms. As estimated by \citet{draine1979}, the thermal sputtering rate for graphite, silicate, or iron grains at rest in hot gas with $10^6 < T < 10^9$ K is of the order of 
\begin{equation}
\tau_{\rm sp} = 2 \times 10^4 \left(\frac{\rm cm^{-3} }{ n_H}\right)\left(\frac{a }{ 0.01 \mu m}\right)\; [{\rm yr}],
\end{equation}
so for the two-temperature dusty region in our computations, the densities are smaller than $\sim$ 1 cm$^{-3}$, and the timescale is of the order of years. However, if strong shocks are present, the dust destruction is more efficient \citep{drainePaperII1979}.

During the photoionization process by intense X-rays near an active nucleus, the dust grains develop an electric charge that is attached to their surface, and thus, a complex dusty plasma is formed \citep{1979ApJ...231...77D,2005pacp.book.....V,2006ApJ...645.1188W,2007JPhD...40..121I}. Collective effects between the dust and gaseous components are essential in this environment, and the dust grain properties are different from those observed in the local interstellar medium. It has been argued that in AGN the dust grains suffer destruction by charging \citep[e.g.][and further references cited therein]{2020ApJ...892...84T}, and the dust can be thus depleted. Also, the equilibrium structure of dusty tori may be significantly altered by the interplay of gravitational and electromagnetic forces that combine near accreting black holes in contrast with the classical test-fluid solutions. In our works \citep{2011PhRvD..84h4002K,2016ApJS..226...12T} we have investigated the emerging structure of the charged tori, which may be quite different from the globally neutral ones. Especially their vertical extent and stability are affected \citep{2020PhRvD.101h3027T}.

\section{Coexistence of hot and warm dusty plasma in the Galactic center}

The Galactic center serves as the nearest to us prototype of a low-luminosity galactic nucleus. The bolometric luminosity of the compact variable source Sgr~A*, which is associated with the supermassive black hole of $M_{\bullet}\sim 4\times 10^6\,M_{\odot}$ \citep{2017FoPh...47..553E,2022RvMP...94b0501G,2022arXiv221107008D}, is $L_{\rm SgrA*}\lesssim  10^{37}\sim 2600\,L_{\odot}\,{\rm erg\,s^{-1}}$ \citep{1998ApJ...492..554N}. The comparison with the Eddington luminosity $L_{\rm Edd}=5.03\times 10^{44}(M_{\bullet}/4\times 10^6\,M_{\odot})\,{\rm erg\,s^{-1}}$ for a given mass yields the Eddington ratio of $\lambda_{\rm Edd}=L_{\rm SgrA*}/L_{\rm Edd}\lesssim 2\times 10^{-8}$. However, the source is variable with order-of-magnitude near-infrared and X-ray flares taking place a few times per day \citep[see, e.g.,][ and references therein]{2021ApJ...917...73W}, which appears to be related to the hot plasma components with the size of $\sim 4$ gravitational radii and the temperature of $\sim 10^{10}\,{\rm K}$, the electron density of $4\times 10^{7}\,{\rm cm^{-3}}$, and the magnetic field of $\sim 10\,{\rm G}$ orbiting close to the innermost stable circular orbit in the form of ``hot spots'' \citep{2018A&A...618L..10G,2022A&A...665L...6W}. There are several signatures that the accretion rate was significantly larger on the timescale of several hundred to million years, which has been revealed by the reprocessing of the central X-ray emission by the surrounding molecular clouds \citep{1993ApJ...407..606S,1998MNRAS.297.1279S} and by the large-scale $\gamma$-ray emitting \textit{Fermi} bubbles and the X-ray emitting \textit{eRosita} bubbles \citep{2010ApJ...724.1044S,2020Natur.588..227P}.

Because of the current low Eddington ratio, the accretion flow surrounding Sgr~A* is hot and diluted with low radiative efficiency. It can be modelled in the framework of advection-dominated accretion flows (or ADAFs) with a large scale-height-to-radius ratio approaching unity. On the larger scale, the flow is Bondi-like with power-law temperature and density profiles of the hot plasma with the asymptotic values of $n\sim 26\,{\rm cm^{-3}}$ and $T\sim 1.5\times 10^7\,{\rm K}$ at the Bondi radius \citep{2003ApJ...591..891B}, which can be estimated as follows,
\begin{align}
    R_{\rm B}&=\frac{2GM_{\bullet}}{c_{\rm s}^2}\simeq \frac{2GM_{\bullet}\mu m_{\rm H}}{kT}\,\notag\\
    &\sim 0.14\,\left(\frac{M_{\bullet}}{4\times 10^6\,M_{\odot}} \right) \left(\frac{T}{1.5\times 10^7\,{\rm K}} \right)^{-1}{\rm pc}\,,
    \label{eq_Bondi_radius}
\end{align}
where $c_{\rm s}$ is the asymptotic sound speed in the plasma. The Bondi radius expressed by Eq.~\eqref{eq_Bondi_radius} corresponds to $R_{\rm B}\sim 7.3\times 10^5$ gravitational radii.
However, the X-ray surface brightness is better described by a flatter density profile than for the standard Bondi inflow, i.e., $n\propto r^{-3/2+s}$, where $s\sim 1$, which implies the presence of an outflow \citep{2013Sci...341..981W}. 

The ADAF is located inside the Bondi radius, which is well within the ionized \textit{central cavity} with the radius of $\sim 1$--$1.5$ parsecs filled mostly with hot X-ray emitting optically thin plasma. In this region, which is also referred to as Sgr A West, there are distinct thermal streamers known as the Minispiral, whose kinematics is consistent with three bundles of Keplerian orbits around Sgr~A* corresponding to the Northern Arm, Eastern Arm, and the Western Arc \citep{2009ApJ...699..186Z, 2010ApJ...723.1097Z}. The Minispiral is well detected in the mid-infrared domain \citep{2022ApJ...929..178B}, which reveals the dust emission, and also in the sub-millimeter, millimeter, and the radio domains \citep{2017A&A...603A..68M,2017ApJ...842...94T}. The electron density of the Minispiral is in the range of $3$--$21\times 10^4\,{\rm cm^{-3}}$ and the electron temperature is $5\,000$--$13\,000\,{\rm K}$ \citep{2010ApJ...723.1097Z,2012A&A...538A.127K}. The highest densities of $21\times 10^4\,{\rm cm^{-3}}$ are coincident with the IRS 13 region, while the highest temperature of $1.3 \times 10^4\,{\rm K}$ is in the Bar region where the Northern and the Eastern Arms appear to collide at $\sim 0.1$--$0.2$ pc south of and behind Sgr~A* \citep{2010ApJ...723.1097Z}. The dust density traces the ionized gas density along the three arms, which leads to the emission enhancement in the mid-infrared domain in the clumps along the Minispiral as well as at the places where supersonic stars flow through the ambient plasma and create denser bow shocks \citep{2002ApJ...575..860T,2003ANS...324..597T,2013A&A...551A..35R,2014A&A...567A..21S}. This shows that dust is generally coupled to gas via gas drag like in the standard interstellar medium, and the dust-to-gas mass ratio of 1:100 \citep{1978ApJ...224..132B} is usually adopted for basic estimates. 

Towards the Galactic center, the near-infrared emission (2.2\,${\rm \mu m}$) is moderately polarized with the mean degree of $\sim 4\%$ and the position angle of $\sim 30^{\circ}$ along the Galactic plane. This foreground polarization is apparently due to anisotropic scattering of infrared radiation on elongated dust grains that are aligned along the Galactic spiral arms between the Earth and the Galactic center \citep{1995ApJ...445L..23E,1999ApJ...523..248O}.  
The presence of dust grains that are dynamically coupled to the sheared gas along the orbiting Minispiral arms can also be revealed via the polarized light. \citet{1991ApJ...380..419A} and \citet{1998MNRAS.299..743A} detected the polarized mid-infrared emission ($12.4\,{\rm \mu m}$) of ionized Minispiral filaments, which corresponds to the thermal emission of elongated dust grains aligned along the magnetic field lines that are approximately parallel to the bulk streaming motion of the Minispiral gas. This indicates that the magnetic field with the strength of $\gtrsim 2$ mG is an inherent property of the sheared ionized gas of the Minispiral. \citet{2018MNRAS.476..235R} confirmed that the mid-infrared polarization is a general property of the diffused ionized gas in the Galactic center. They found the highest polarization degree of $\sim 12\%$ at 12.5 ${\rm \mu m}$ in the Northern Arm. The polarized emission of embedded bow-shock sources is generally enhanced due to the compressed magnetic field, and the polarization degree also increases from near-infrared to mid-infrared domains, which implies that thermal emission by aligned dust grains, i.e., the reprocessed UV emission of the central star, is responsible for the polarization properties \citep{2011A&A...534A.117B, 2013A&A...557A..82B}. For the most prominent bow-shock sources - such as IRS 21 and IRS 1W, the degree of polarization of the integrated infrared emission falls in the interval of $\sim 10$--$30\%$. This can be explained by the Mie-scattering on dust grains, whose density is enhanced in asymmetric gaseous-dusty shells surrounding the supersonic stars \citep{1995ApJ...445L..23E,1999ApJ...523..248O}.   

The ionized \textit{central cavity} ends quite sharply at $\sim 1.5$ parsecs. Its radius is determined by the ambient UV field of the nuclear star cluster, which is concentrated in the inner parsec. Further away, from about 1.5 pc to $\sim 3$--$4$ parsecs, a dense ring-like circumnuclear disk (CND) is located that is composed of neutral and molecular gas and dust with the equilibrium temperature between 20 and 80 K \citep{1982ApJ...258..135B, 1989A&A...209..337M}. The gas number density falls into the range $\sim 10^5$--$10^8\,{\rm cm^{-3}}$, being the densest in numerous clumps along the ring \citep{2005ApJ...622..346C} that have typical sizes of $\sim 0.25\,{\rm pc}$ and masses of a few $10^4\,M_{\odot}$. The whole CND has a mass of $\sim 10^6\,M_{\odot}$, being about three to four orders of magnitude more massive than the Minispiral, whose ionized mass is $\sim 350\,M_{\odot}$ \citep{2010ApJ...723.1097Z}. Because of its higher mass and several tens of dense clumps, it has the potential to provide the material for the in-situ star formation in the nuclear star cluster \citep{2018ApJ...864...17T}, though there is no evidence of ongoing star formation; most likely turbulence and magnetic field prevent most clumps from collapsing to form stars \citep[see, e.g.,][]{2021ApJ...913...94H}. The CND and the Minispiral streamers are, interestingly, not coplanar. While the Northern Arm and the Western Arc reside in nearly the same plane, which seems to be coincident with the plane of the CND, the Eastern Arm is nearly perpendicular to the \citep{2000NewA....4..581V,2010ApJ...723.1097Z}. In this regard, the Northern Arm and the Western Arc might have originated in the inner rim of the CND, where they got detached by the combined effect of photoionization and the interaction with the  outflow from the nuclear star cluster. The Minispiral is, therefore, a transient, disappearing feature with a dynamical timescale of $\sim 10^4$ years. At larger distances from the Sgr A complex, there are several massive molecular clouds, such as ``$+20\,{\rm km\,s^{-1}}$'' and ``$+50\,{\rm km\,s^{-1}}$'' giant molecular clouds \citep[see][for a review]{1996A&ARv...7..289M}.   

\begin{figure}
    \centering
    \includegraphics[width=0.8\textwidth]{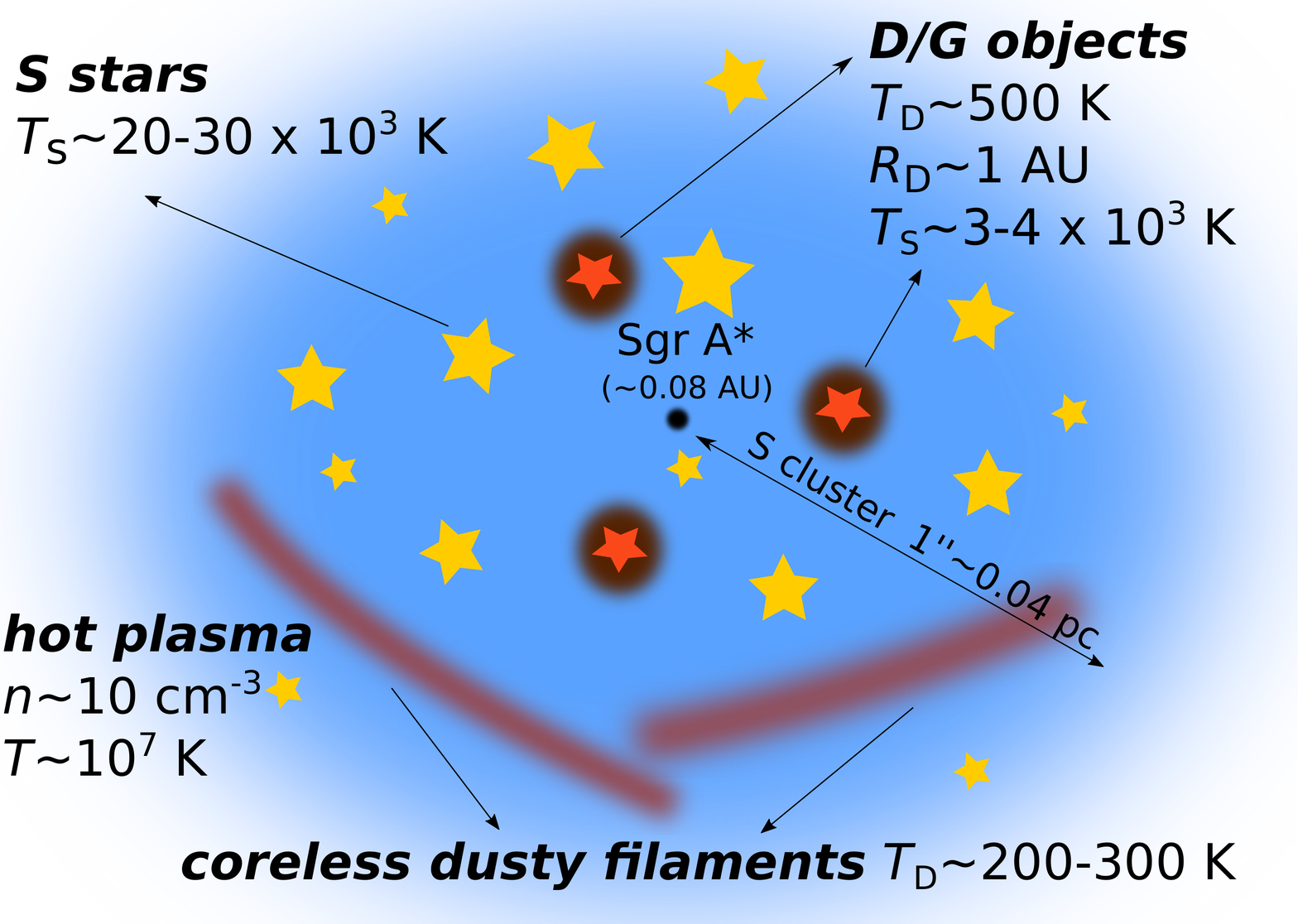}
    \caption{Illustration of the central region of the Milky Way (S-cluster, $1^{\prime\prime}\sim 0.04$~pc). Dust-embedded D/G sources are characterized by a higher dust temperature of $\sim 500$ K with the photospheric radius of $R_{\rm D}\sim 1$ AU, while the coreless Minispiral streamers have a dust temperature of $\sim 200$--$300$ K. The main-sequence stars within the S-cluster are predominantly of spectral type B with the effective temperature of $20$--$30\times 10^3$ K. The entire region of the Galactic center is filled with optically thin, X-ray emitting plasma with the number density of $n\sim 10\,{\rm cm^{-3}}$ and $T\sim 10^7$ (the figure redrawn from \citealt{2020A&A...634A..35P}).}
    \label{fig_dusty_sources}
\end{figure}

\begin{figure}
    \centering
    \includegraphics[width=\textwidth]{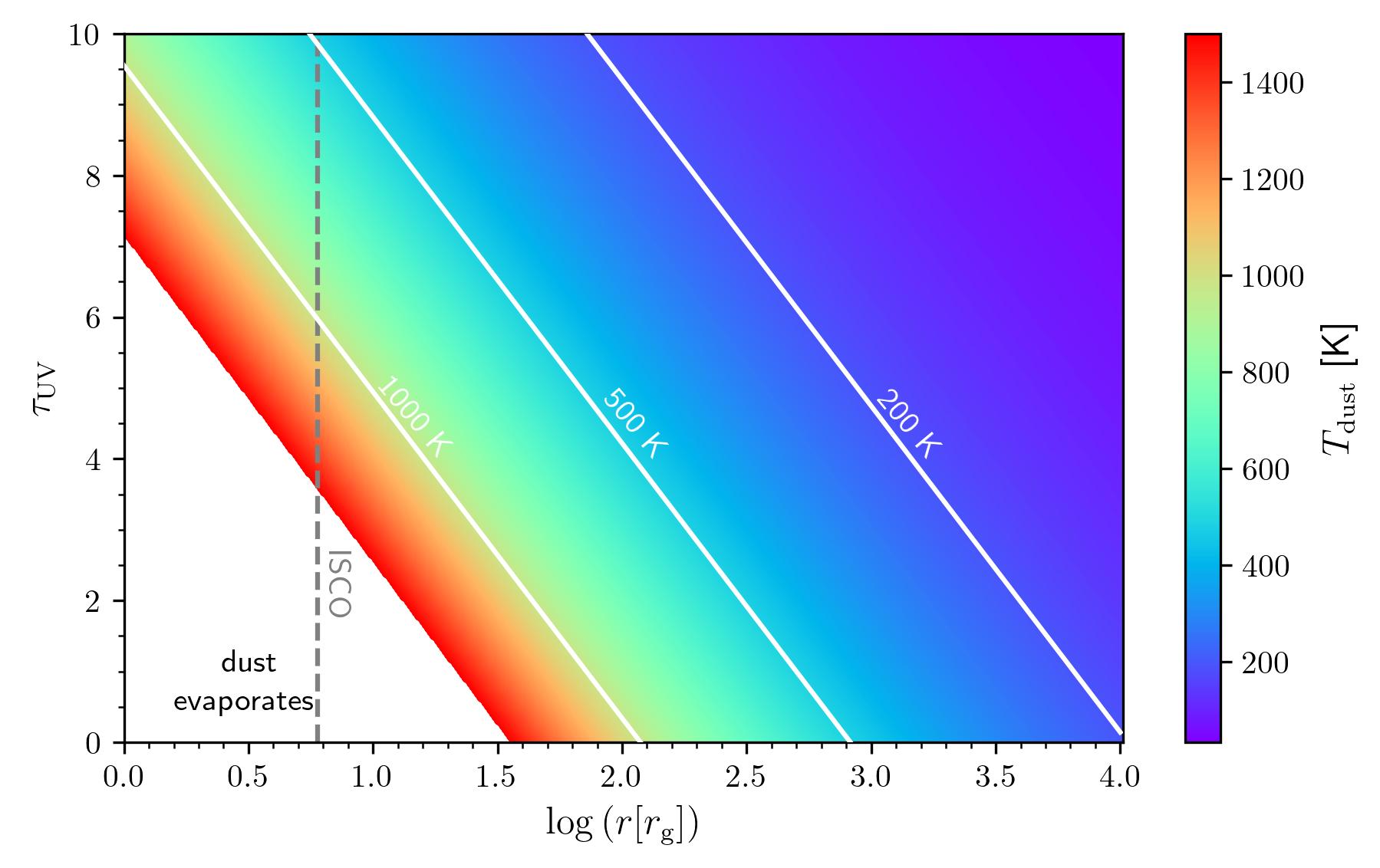}
    \caption{Colour-coded dust temperature distribution as a function of the distance $r$ (in gravitational radii) from a low-luminosity galactic nucleus and the UV optical depth $\tau_{\rm UV}$. The calculation considers the central source UV luminosity of $L_{\rm UV}=100\,L_{\odot}$. The white region depicts the dust evaporation zone. The gray dashed vertical line stands for the location of the innermost stable circular orbit (ISCO) around a Schwarzschild black hole ($6\,r_{\rm g}$). White solid lines represent the dust temperature of 200, 500, and 1000 K.}
    \label{fig_dust_temp}
\end{figure}

In the central cavity, dust is present and observed, not only within the Minispiral but also closer in within the S cluster, i.e., within $1^{\prime\prime}\sim 0.04\,{\rm pc}$ from Sgr~A*. Coreless gaseous-dusty filaments within the Minispiral are characterized by warm dust with the temperature of $\sim 190$--$265$ K, which was inferred based on the 12.5/20.3 ${\rm \mu m}$ color-temperature map \citep{1999ASPC..186..240C}. In the S cluster, fast and compact dusty D or G objects with emission lines of hydrogen and helium have been identified \citep{2012Natur.481...51G,2013A&A...551A..18E,2020A&A...634A..35P,2020Natur.577..337C,2021ApJ...923...69P} that orbit Sgr~A* on eccentric orbits similar to S stars. They are characterized by the optically thick envelopes with the photospheric radius of $\sim 1\,{\rm AU}$ and the black-body temperature of $\sim 500$--$600$ K \citep{2020A&A...634A..35P}, which can be attributed to the dust thermal emission. The higher temperature of dust in comparison with dusty streamers can be explained by the model of a dust-enshrouded star \citep{2014A&A...565A..17Z, 2017A&A...602A.121Z}, i.e., an optically thick dusty shell is heated by a central star, see Fig.~\ref{fig_dusty_sources} for an illustration. Although the nature of D/G sources is disputed, the pericenter passages of G2/DSO and G1 sources can be used to put an upper limit on the dust sublimation radius around Sgr~A*, $R_{\rm sub}\sim 137\,{\rm AU}=3469\,r_{\rm g}$ and $R_{\rm sub}\sim 298\,{\rm AU}=7577\,r_{\rm g}$ for G2/DSO and G1 pericenter distances \citep{2017ApJ...847...80W,2021ApJ...923...69P}, respectively. Because of the current low-luminous state of Sgr~A*, it can be shown that dust can approach Sgr~A* within the innermost 100 gravitational radii if it is not destroyed before in shocks and by the UV emission of numerous OB stars. However, it cannot approach or form and linger as close as the innermost stable circular orbit if it is not shielded. We estimate the corresponding sublimation radius $r_{\rm sub}$ for graphite dust following \citet{barvainis1987}, assuming the sublimation temperature of $T_{\rm sub}\approx 1500\,{\rm K}$, the zero optical depth $\tau_{\rm UV}=0$, and the UV luminosity of Sgr~A* with the upper limit given by its current bolometric luminosity $L_{\rm UV}\lesssim L_{\rm bol}\approx 100\,L_{\odot}$ \citep{2001ARA&A..39..309M}. The sublimation radius scaled to the gravitational radius can then be expressed as
\begin{equation}
    \frac{r_{\rm sub}}{r_{\rm g}}=38.07 \left[\left(\frac{L_{\rm UV}}{100\,L_{\odot}}\right) \exp{(-\tau_{\rm UV})} \right]^{\frac{1}{2}} \left(\frac{T_{\rm sub}}{1500\,{\rm K}} \right)^{-2.8}\left(\frac{M_{\bullet}}{4\times 10^6\,M_{\odot}} \right)^{-1}\,.
    \label{eq_rsub}
\end{equation}
The distribution of dust temperature $T_{\rm dust}$ up to the sublimation temperature of $1500$ K as a function of the distance from Sgr~A* (in gravitational radii) is plotted in Fig.~\ref{fig_dust_temp}, following the simple model of \citet{barvainis1987} that takes into account only the central luminosity source. We consider the optical depth in the range $\tau_{\rm UV}\in (0,10)$ and the UV luminosity of Sgr~A* $L_{\rm UV}=100\,L_{\odot}$, which is approximately achieved during daily infrared/X-ray flares \citep[see e.g.][]{2014ARA&A..52..529Y}. It is clear that dust cannot exist close to the innermost stable orbit (dashed gray vertical line for a non-rotating black hole; $6\,r_{\rm g}$) unless it is shielded by an optically thick envelope. Conditions for the survival of the clouds drifting in/out or forming in changing conditions are more complex since then the cooling timescales must be included \citep[e.g.][]{rozanska2014,rozanska2017,2022ApJ...929..178B}.

Overall, the Galactic center stands for a unique environment where in the gravitational sphere of influence of the supermassive black hole ($\sim 2$ pc) one can detect cold and dense molecular gas ($T\sim 50$ K and $n\sim 10^8\,{\rm cm^{-3}}$) as well as plasma close to Sgr~A* that is dense and hot ($T\sim 10^{10}\,{\rm K}$ and $n\sim 10^7\,{\rm cm^{-3}}$). At intermediate scales of $\sim 0.1$--$1\,{\rm pc}$, warm ($\sim 10^4$ K) and hot gas phases ($\sim 10^7$ K) coexist in the Minispiral region alongside the warm dust with the temperature of a few 100 K. Unfortunately, we do not have this high-resolution multi-wavelength images  of other AGN.

\section{Nuclear dust properties in AGN from spectral and time delay studies}
\label{sect:dust_properties}

In highly accreting objects, the total emission is dominated by the optical/UV radiation from the cold disk, and there is an additional bump in the near-IR \citep{neugebauer1979}
which comes from the dust reprocessing \citep{barvainis1987,pier1992,fritz2006}. The basic geometry was set by the discovery of \citet{antonucci1985}, but subsequent spectral studies shed more light onto the dust content, including the chemical composition. 

The properties of dust are the subject of vigorous studies for a long time. The dust chemistry is very complex \citep[see][for reviews]{laor_draine1993,li2007,ramosAlmeida2017}.
Observations reveal spectral (mostly silicate) features in emission \citep[e.g.][]{siebenmorgen2005,hatziminoglou2015} and in absorption \citep[e.g.][]{aitken1985,roche2007,tsuchikawa2021}, depending on the viewing angle in each specific source. Polycyclic Aromatic Hydrocarbons (PAH), well observed in the interstellar medium are frequently weak or not visible in AGN since they are most likely destroyed by shocks \citep[e.g][]{zhang2022,xie2022}. The extinction curve that should be used to correct the intrinsic spectra of AGN with a certain level of extinction is still an open issue. The standard extinction curve from the Milky Way \citep{cardelli1989} cannot be applied since the 2175 \AA\ feature is generally not observed, but various specific shapes were proposed \citep[e.g.][]{czerny2004,gaskell2004,gaskell2022}. This would favor amorphous carbon grains. In some objects, however, the 2175 \AA~ feature seems to be present \citep[][in FeLoBAL objects]{zhang_dust2022}.

As discussed in Sect.~\ref{sect:basic}, dust has to be concentrated in some form of a torus-like geometrical structure that is responsible for obscuring the inner parts of the nucleus in type 2 sources, while not being an obstacle blocking the view of the black hole region in type 1 sources (despite that it contributes to the IR band). Statistically, the fraction of the sky covered by the torus is thus responsible for the ratio of type 1 to type 2 sources. In a given source, the ratio of the UV emission to the IR emission also contains information about the angular extension of the dust or the covering factor \citep[e.g.][]{lawrence2010,gu2013}.   

Some constraints for the dust location came first from the simple consideration of the dust sublimation temperature \citep{barvainis1987,nenkova2008a}. The subsequent reverberation measurements (RM) of IR emission time delays  with respect to the optical emission gave more specific constraints \citep[e.g.][]{okny1999,koshida2014,guise2022}. The distance of the dusty emitters from the central black hole is effectively by a factor of 5 larger than the distance of the BLR, as determined from RM. The dust distance in a given source responds to the changes of the radiation flux, and as suggested by the data for the source NGC 4151, the recovery of the dust after the episode of a high luminosity state of the source takes a few years \citep{okny2008}.

A novel and interesting method to localize the dust region in the equatorial plane was proposed by \citet{shablovinskaya2020}. The method is based on reverberation measurement of the broad emission lines in polarized light in unobscured objects. For the studied source Mrk 6, they found that the size of the scattering region is around 100 light days. That is significantly smaller than the dusty region size estimated by the infrared interferometric observations. 

The dusty torus localized around the equatorial plane is a key element of the AGN unification scheme, as mentioned in Section~\ref{sect:basic}. Its evolution with redshift is the main element in cosmological applications, as the average viewing angle of AGN depends on the average opening angle of the shielding torus \citep[see, e.g., the discussion in][]{prince2022}. It is also important for all population studies of AGN. Therefore, the measurements of the correlations of the torus covering factor (or opening angle) are pursued. It seems that, for AGN with the bolometric luminosity larger than $10^{44}$ erg s$^{-1}$, no trends are seen with the luminosity or redshift \citep[e.g.][]{netzerLani2016,stalevski2016,prince2022}. At lower luminosities, the torus seems to disappear, as discussed in Section~\ref{sect:basic}.

\section{Problems with the standard torus model and development of the most recent models}
\label{sect:recent_models}

The simple idea of an AGN torus, as originally introduced by \citet{antonucci1985} and presented in numerous reviews of AGN structure, geometrical relations, and physical properties \citep[e.g.][]{urry1995} is naturally a simplification. Firstly, a stationary torus shape cannot be supported just by thermal motions. It needs some sort of dynamical structure (inflow or outflow); a wind outflow seems to be the most likely option, and it was elaborated in numerous papers \citep[e.g.][]{Konigl1994, elitzur2006,dorodnitsyn2008,dorodnitsyn2021, gallagher2015}. As we now know from detailed spectral modelling, the torus has to be clumpy; otherwise, the broad band spectrum and the absorption/emission features cannot be explained \citep[][]{nenkova2008a,nenkova2008b,stalevski2012,siebenmorgen2015}. 

High-resolution spatial observations reveal another problem. The original expectations were such that the high-resolution IR maps will show predominantly the dusty/molecular torus. However, the images brought us a surprise: a lot of dust emission actually came from the structure elongated in the direction of the symmetry axis. For example, \citet{Seba2013} performed interferometric mapping of the
unobscured (type 1) AGN in NGC 3783. Fitting the visibility plane, they have reached the following conclusion: most of the (mid-IR) emission comes from the polar direction. Specifically, the ratio of the polar to equatorial emission in this source turned out to be a rising function of the wavelength. The polar scattering region was subsequently revealed in a number of single-dish detections \citep[e.g.][]{asmus2016,asmus2019} as well as interferometric observations \citep[][]{lopez2016}. Such observations have led to a picture of the dust distribution in a form of an empty cone suggested by \citet{Seba2019}. In the cross-section, it looks like a narrow stream of escaping dusty material, with a launching radius at the usual location of the beginning of the torus, i.e., about 1 pc from the black hole. Recently developed codes for the radiative transfer combine the disk and polar emitters. These are able to explain the observational data very well \citep[e.g.][]{stalevski2017,stalevski2019}. Moreover, a clumpy disk with the wind models provides the best fits for the nuclear IR spectral energy distributions of partially obscured Seyfert galaxies \citep{garcia2022}.

Present-day advanced computations include the hydrodynamical simulations combined with the effects of radiation pressure. These models reveal the complexity of the medium, not just the dynamics of the outflow but also the role of convection and of complex vertical stratification \citep[e.g.][]{dorodnitsyn2012,dorodnitsyn2021}. Magnetic fields can also play a significant role \citep{dorodnitsyn2017}, as well as the torus self-gravity \citep{1989ApJ...341..685S,2004CQGra..21R...1K}.

The outer disk/torus eventually overlaps with the interstellar medium and possibly also stars that naturally belong to the circumnuclear environment. Star formation in the outer parts of self-gravitating disks leads to supernovae eruptions that are capable of launching strong winds, as proposed by \citet{collin1999}. The medium there is clearly clumpy and dynamic. Only the most recent hydrodynamical models start to address this issue in a realistic manner \citep[e.g.][]{wada2016,williamson2020}. Such outflows clearly affect the host beyond the central parsec, and they enrich the interstellar medium with heavy elements. 

\section{Dust within the Broad Line Region and the FRADO model}

The standard unification scheme locates the dust much farther away than the BLR. However, the possible presence of dust in the broad-line regions of AGN has been debated for many years \citep{goodrich1995}. It is well-known that the ratios of the broad hydrogen Balmer emission lines are not consistent with simple predictions, and the potential presence of dust can contribute to the resolution of this issue \citep[e.g.][and the references therein]{ilic2012}. Medium fully exposed to the central flux cannot retain dust at a distance smaller than the torus inner radius, but the disk in AGN is geometrically thin, and its inner part -- or winds -- may, in addition, shield the disk. This would allow for the dust survival close in.  

Interesting measurements of the time lag between the non-polarized continuum and the polarized broad line variability in the Mrk 6 implied the presence of a scatterer residing much closer than the torus but further away than the BLR \citep{shablovinska2020}. 

It is also a matter of fact that broad lines appear in coexistence with a cold disk located much more centrally than the torus, where the effective temperature of the accretion disk drops below the order of 1000 K, which implies the formation of low ionized dusty medium \citep[see][]{czerny2004,loska2004}. This is the basis of a model developed by \cite{czerny2011} based on the dust-driving mechanism in which the radiation pressure of the disk is responsible for the formation of a low-ionized BLR due to the radiative acceleration of dust. The model predicts the formation of failed winds through frequent elevation of the dusty environs, illuminated by the intense central disk radiation. The subsequent fallback of dustless material occurs onto the disk: Failed Radiatively Accelerated Dusty Outflow (FRADO). The dust-driving mechanism provides an attractive, theory-motivated explanation of a natural source of low ionized broad emission lines. The scheme is supported by the measured time delays \citep{wandel1999, peterson1999, grier2013, zajacek2021, 2021A&A...650A.154P}. These authors have shown that Low Ionization Lines (LIL), such as H$\beta$ and MgII, originate from a cooler and denser medium located farther with respect to the central black hole compared to the other population of broad emission lines, i.e. High Ionization Lines (HIL) such as CIV and HeII for which the line-driving mechanism is favored \citep{murray1995, proga2000, waters2021}.

The new advanced numerical 2.5D FRADO computations opened up a window to the dynamical modeling of low-ionized sections of BLR. Firstly, preliminary tests showed that the 2.5D model can explain the observed location and dispersion in the radius-luminosity relation based on the reverberation-known position of the LIL BLR \citep{peterson2004, bentz2009, bentz2013}, on the basis of accretion rate \citep{naddaf2020}. Moreover, the 2.5D FRADO pictures a rather complex pattern of motions \citep{naddaf2021, naddaf2022}, strongly depending on the global parameters: the accretion rate, the black hole mass, and also the metallicity of material. For small values of black hole mass and accretion rate, this pattern resembles the model of a static, puffed-up disk \citep{baskin2018} with some level of turbulence. Otherwise, an outflow structure forms similar to the one proposed by \citet{elvis2000} and \citet{Seba2019}. Metallicity amplifies the radiative force and intensifies the outflow for the same parameters of accretion rate and black hole mass. This mechanism manifests itself in the shape of broad emission lines \citep{naddaf2022}.

\citet{Seba2019} proposes an outflow structure initiated at radii typical for the onset of the torus (of around 10--100 times the predictions of 2.5D FRADO), thus representing the torus as a separate nested structure. On the other hand, there are studies indicating a likely connection between the BLR and the torus. The ample amount of dust of high column density extending from the equatorial plane to high altitudes and obscuring the central disk at high inclinations \citep{antonucci1985} implies a very dynamical torus structure, as already mentioned in Sect.~\ref{sect:recent_models}. Moreover, the systematic study of the absorption events caused by the dusty clouds \citep{markowitz2014} may entail  the BLR and torus as overlapping structures. A combination of the above-mentioned hints suggests that the BLR and the dusty torus are closely interrelated rather than separate identities \citep{wang2013}. Last but not least, the dust geometry proposed recently on the basis of the observational data \citep[e.g.][]{elvis2000, kawaguchi2010, kawaguchi2011, goad2012, figaredo2020} is consistent with the outflow structure in the theoretically motivated 2.5D FRADO scheme \citep{naddaf2021, naddaf2022}. 

\section{Dust scenario in Changing Look AGN}

In the standard unification scheme, the AGN type is determined for a given source by its viewing angle; therefore, it should not vary. Also, in X-rays, the amount of obscuration towards the nucleus in a given source should be set by geometry and remain unchanged. However, examples of drastic changes of the X-ray obscuration \citep{matt2003} and of appearance and disappearance of the broad lines in some sources were reported \citep{alloin1985}. Such objects are now referred to as Changing Look AGN (CL AGN) phenomenon. The effect was considered rather rare in the past, however, as a result of follow-up detections the sample has recently grown \cite[e.g.,][]{Yang2018, graham2020, panda2022}. 

Two different physical scenarios have been tested to explain spectral and luminosity changes in CL AGN: the obscuration mechanism and the variation in the accretion state. In fact, most of the changing-look phenomena seem to be caused by accretion variation rather than a variable obscurer. The temporary obscuration or intrinsic dimming scenarios were also tested using spectropolarimetry since a low level of polarization argues against the obscuration scenario \citep{2019A&A...625A..54H}. \citet{lamassa2015} discussed different obscuration scenarios for CL AGN SDSS J015957.64+003310.5., which was reported as the first CL QSO. Tentative obscuration scenarios consider the dust  transported from the area of the torus that could cover the BLR emission or the processes of enhanced dust formation. However, for both scenarios, the timescales are longer than the observed changes. Timescales disagreement is the most robust argument for most CL AGNs against the obscuration scenario. The possible presence of the dust within BLR and the atypical extinction properties of AGN dust may open some paths to use obscuration as a plausible CL mechanism. In the case of intrinsic changes, timescales are also much shorter than normal viscous timescales in the standard accretion disks, and appropriate model modifications have to be made to accommodate the observational data \citep[e.g.][]{sniegowska2020,sniegowska2022,kaur2022,pan2022}. Further studies are clearly needed.

\section{Dust origin in AGN}

Dust in the BLR and in the dusty/molecular torus can either come from the interstellar medium, or it can form {\em in situ}. The correct answer is currently unclear. The solar and frequently super-solar metallicity, even at high redshift quasars, is not consistent with the expected low metallicity of the high redshift galaxies \citep[see, e.g.,][and further references therein]{sniegowska2021,panda2021}. On the other hand, vigorous star formation actually precedes the quasar activity, and multiple supernovae in the circumnuclear stellar cluster creating heavy elements enrich the material that later accretes  \citep[see][for the most recent plot of the history of star formation rate and of quasar activity]{fiore2017}. 

The  {\em in situ} option is interesting. Outer parts of the disk can be gravitationally unstable, leading to vigorous star-forming, with predominantly fast-evolving massive stars \citep[e.g.][]{jermyn2021,cantiello2021}. What is more, the standard disk in the region of the effective temperature $\sim$ 1500 K allows the formation of dust directly in the disk atmosphere, in a similar way as in stars, and this mechanism was already discussed by \citet{elvis2002}. 

Local production can possibly supply the dust much faster than the external source if dust is temporarily destroyed during the bright stage of an AGN, although more studies of this issue similar to the tentative reports by \citet{okny2008} are needed.

\section{Summary}
\label{sect:summary}

The presence of the dust in AGN is well known from the spectroscopic studies done back in the 1960s and 1970s \citep[see ][and the references therein]{osterbrock1979}. Over the years, there has been gradual progress in our knowledge of the dust location and distribution that came from polarimetric studies, interferometric measurements in the IR, and time-domain measurements. We understand the role of dust in the AGN classification scheme, and we have an insight into the dynamics of this complex, clumpy region. However, there are still important unresolved questions related to the formation of the dusty region and its connection with the host galaxy. 

The most important problem is the relation of the dust to the accretion disk. It may seem that the absence of the cold disk and the presence of the exclusively hot flow prevents the existence of the BLR and of the dusty torus. However, such a hypothesis is not proven observationally, and it is not easy to test it since, at a fraction of a parsec, the presence of the disk overlapping with BLR and torus region is hard to resolve. The polarization-based arguments of the decomposition of quasar spectra support the idea of the cold disk underlying the two regions \citep{kishimoto2008}. The progress in IR interferometric studies will certainly shed light on the exact geometry of the dust close to the dust sublimation radius \citep[see, e.g.,][for one of the most recent results for NGC 4151]{kishimoto2022}. Such values done for more objects, and possibly in the same object repeated after a few years, can help us to understand the evolutionary aspects.

Interesting new developments are now coming from the James Webb Space Telescope (JWST) launched at the end of 2021. Indeed, one of the four general aims of JWST is to trace the process of the assembly of galaxies. Since AGN is an important stage in a galaxy's life, it will also shed light on the growth of the central black hole and the feeding and feedback. Dusty outflows in AGN are massive so they are an important element of the overall puzzle. We thus, expect a lot of new developments in dust formation and dust circulation in galaxies, including their AGN phase. Infrared observations give a direct insight into star formation in dusty galaxies, including the interacting galaxies \citep[e.g.][]{evans2022}. Particular attention is paid to extremely red quasars, with bolometric luminosity exceeding $10^{47}$ erg s$^{-1}$ at high redshift \citep[e.g.][]{wylezalek2022}. Sources like the one studied in this paper (SDSS J165202.64+172852.3) reveal the formation stage of the core of the clusters of galaxies. These are excellent candidates for super-Eddington accretion important to explain the fast rise of the black hole mass from the initial seeds that are required to match the observed distribution of the black hole masses \citep[][]{2020ARA&A..58...27I,2023MNRAS.519.4753T}. It also allows to map of the inner $\sim 1$ kpc of nearby AGN with the resolution of $\sim 100 $ pc and sheds light on the circumnuclear starburst ring and the associated outflow \citep[e.g.][]{U2022}. Such observations allow us to quantify the vigorously discussed issue of the feedback of an active nucleus to the host as well as the feeding pattern of the central black hole.




\section*{Acknowledgments}
This project has received funding from the European Research Council (ERC) under the European Union’s Horizon 2020 research and innovation program (grant agreement No. [951549]). The project was partially supported by the Polish Funding Agency National Science Centre, project 2017/26/A/ST9/00756 (MAESTRO 9), and MNiSW grant DIR/WK/2018/12. MLM-A acknowledges financial support from Millenium Nucleus NCN$19\_058$ (TITANs). VK acknowledges the Czech Science Foundation project No. 21-06825X ``Accreting black holes in the new era of X-ray polarimetry missions''. SP acknowledges financial support from the Conselho Nacional de Desenvolvimento Científico e Tecnológico (CNPq) Fellowship (164753/2020-6). MS acknowledges the financial support of the Polish Funding Agency National Science Centre 2021/41/N/ST9/02280 (PRELUDIUM 20). MZ acknowledges the financial support of the GA\v{C}R EXPRO grant No. 21-13491X ``Exploring the Hot Universe and Understanding Cosmic Feedback''. 
The authors also acknowledge the Czech-Polish mobility program (M\v{S}MT 8J20PL037 and
NAWA PPN/BCZ/2019/1/00069), the OPUS-LAP/GA\v{C}R-LA bilateral project (UMO-2021/43/I/ST9/01352 and GF22-04053L), and the Czech Ministry of Education, Youth and Sports Research Infrastructure (LM2023047). 

\section*{Declarations}

\begin{itemize}
\item Funding

The funding details are listed in the Acknowledgement section.

\item Conflict of interest/Competing interests (check journal-specific guidelines for which heading to use)

The authors declare they have no financial interests.

\item Ethics approval 

Not applicable

\item Consent to participate

All authors contributed to the work and approved sending the paper for publication.

\item Consent for publication

All authors agree for the work to be published in the European Physical Journal D,
Topical Issue: "Physics of Ionized Gases and Spectroscopy of Isolated Complex
Systems: Fundamentals and Applications".

\item Availability of data and materials

Data can be available on request.

\item Code availability 

Codes can be available on request.

\item Authors' contributions

All authors contributed to the study's conception and design. B.C. was the leading author in writing the text.  M.Z., M.H.N., M.S., S.P. and V.K. wrote some parts of the text. T.P.A. provided a better version of Fig. 1. S.P. and V.K.J. helped considerably with text formatting. M.S. and M.L.M.-A. provided a number of references.  All authors sent their comments to the first draft.  Finally, all authors carefully reviewed and approved the final version of the text.
\end{itemize}







\bibliography{sn-bibliography}

\end{document}